\documentclass{aa}  

\usepackage{graphicx}
\bibpunct{(}{)}{;}{a}{}{,}

\usepackage{txfonts}
\usepackage{amssymb}
\begin{document}

   \title{Small vs large dust grains in transitional disks: \\ do different cavity sizes indicate a planet?}

   \subtitle{SAO 206462 (HD 135344B) in polarized light with VLT/NACO\thanks{Based on observations collected at the European Organisation for Astronomical Research in the Southern Hemisphere, Chile, under program number 089.C-0611(A).} }

   \author{A. Garufi
          \inst{1}
          \and
          S.P. Quanz
          \inst{1}
          \and
          H. Avenhaus
          \inst{1}
          \and
          E. Buenzli
          \inst{2, 3}
          \and
          C. Dominik
          \inst{4}
          \and
          F. Meru
          \inst{1}
          \and \\
          M.R. Meyer
          \inst{1}  
          \and
          P. Pinilla
          \inst{5, 6}
          \and
          H.M. Schmid
          \inst{1}
          \and
          S. Wolf
          \inst{7}
          }

   \institute{Institute for Astronomy, ETH Zurich, Wolfgang-Pauli-Strasse 27, CH-8093 Zurich, Switzerland\\
              \email{antonio.garufi@phys.ethz.ch}
         \and
             Department of Astronomy and Steward Observatory, University of Arizona, Tucson, AZ 85721, USA
         \and
             Max-Planck Institute for Astronomy, K\"{o}nigstuhl 17, D-69117, Heidelberg, Germany
         \and
         Sterrenkundig Instituut Anton Pannekoek, Science Park 904, 1098 XH Amsterdam, The Netherlands
         \and
         Universit\"{a}t Heidelberg, Zentrum f\"{u}r Astronomie, Institut f\"{u}r Theoretische Astrophysik, Albert-Ueberle-Str. 2, 69120 Heidelberg, Germany
         \and
         Leiden Observatory, Leiden University, PO Box 9513, 2300 RA Leiden, the Netherlands
         \and
         University of Kiel, Institute of Theoretical Physics and Astrophysics, Leibnizstrasse 15, 24098 Kiel, Germany  
             }

   \date{Received ...; accepted ...}

 
  \abstract
   {Transitional disks represent a short stage of the evolution of circumstellar material. Studies of dust grains in these objects can provide pivotal information on the mechanisms of planet formation. Dissimilarities in the spatial distribution of small ($\mu$m$-$size) and large (mm$-$size) dust grains have recently been pointed out.}
   {Constraints on the small dust grains can be obtained by imaging the {distribution} of scattered light at near-infrared wavelengths. We aim at resolving structures in the surface layer of transitional disks (with particular emphasis on the inner $10-50$~AU), thus increasing the scarce sample of high-resolution images of these objects.}
   {We obtained VLT/NACO near-IR high-resolution polarimetric differential imaging observations of SAO 206462 (${\rm HD 135344B}$). This technique allows one to image the polarized scattered light from the disk without any occulting mask and to reach an inner working angle of $\sim 0.1\arcsec$.}
   {A face-on disk is detected in $H$ and $K_{\rm s}$ bands between $0.1\arcsec$ and $0.9\arcsec$. No significant differences are seen between the $H$ and $K_{\rm s}$ images. {In addition to the spiral arms, these new data allow us to resolve for the first time an inner disk cavity for small dust grains. The cavity size ($\simeq 28$ AU) is} much smaller than what is inferred for large dust grains from (sub-)mm observations {(39 to 50~AU). This discrepancy cannot be ascribed to any resolution effect.} }
   {The interaction between the disk and potential orbiting companion(s) can explain both the spiral arm structure and the discrepant cavity sizes for small and large dust grains. One planet may be carving out the gas (and, thus, the small grains) at 28~AU, and generating a pressure bump at larger radii (39~AU), which holds back the large grains. We analytically estimate that, in this scenario, a single giant planet (with a mass between 5 and 15~${\rm M_J}$) at 17 to 20~AU from the star is consistent with the observed cavity sizes.}


\authorrunning{Garufi et al.}

\titlerunning{Small vs large dust grains in transitional disks}

   \maketitle
%

\section{Introduction} \label{Introduction}
To enhance our understanding of planet formation, a study of circumstellar material at different evolutionary stages is necessary. Most young stars (with ages up to a few Myr) harbor circumstellar disks which are thought to be the birthplaces of planets. A small sample of disks, the so-called transitional disks, shows a peculiar dip in the infrared spectral energy distribution (SED) indicating a depletion of warm dust near the central star \citep{Strom1989, Skrutskie1990}. Since this {type} of disks represents only a few percent of the total, the process of dispersal of disk material must occur rapidly \citep[less than 1 Myr, ][]{Muzerolle2010}. Mechanisms responsible for such a fast dissipation are still under debate. Gravitational interaction with an orbiting companion \citep[e.g.][]{Rice2003, Ireland2008}, photoevaporation \citep{Alexander2007}, and particle growth \citep{Dullemond2005} have been proposed as clearing processes and {innumerable attempts} are made to predict the rapidly increasing observational evidence. 

One of the proposed mechanisms of planet formation is dust coagulation in the inner few tens of AU \citep[e.g.][]{Dullemond2004}. Direct imaging of those regions is therefore essential. Polarimetric differential imaging (PDI) is a high-contrast technique allowing imaging of near-IR scattered light from circumstellar disks with unprecedented inner working angle \citep[e.g.][]{Apai2004, Quanz2011, Quanz2012, Quanz2013, Hashimoto2012}. This technique takes advantage of the fact that direct stellar light is essentially unpolarized whereas scattered light from dust grains from the disk surface is polarized{. PDI} allows direct imaging of regions as close as $0.1\arcsec$ to the central star.  

The study of different dust components in and around the central cavity of transitional disks provides insights into the mechanisms of planet formation. Direct detections of these cavities have been provided, among others, by the Submillimeter Array (SMA) interferometer \citep[e.g.][]{Andrews2009, Andrews2011} at sub-millimeter wavelengths, and by VLT/NACO \citep[e.g.][]{Quanz2011, Quanz2013} and Subaru/HiCiao \citep[e.g.][]{Mayama2012, Hashimoto2012} in near-IR polarized light. Sub-millimeter and near-IR data probe different dust grain sizes (and, thus, different disk layers). Intriguingly, in some cases they are suggesting unexpected differences in the small and large dust grain distribution. \citet{Dong2012} highlighted that the polarized near-IR images from Subaru/HiCiao do not always show the inner cavities observed at millimeter wavelengths with SMA as close as the inner working angle ($0.1\arcsec - 0.2\arcsec$). Their model suggests a decoupled spatial distribution of the big and small dust particles inside the cavity.

In this paper, we present PDI observations of the transitional disk around SAO 206462 (HD135344B) obtained with VLT/NACO. SAO 206462 is an extensively studied, rapidly-rotating \citep{Mueller2011} {F4Ve star \citep{Dunkin1997}} hosting a transitional disk. It is located in the Sco OB2-3 star-forming region \citep[$d=140 \pm 42$ pc,][]{vanBoekel2005}. The angular separation from HD 135344A is 21$\arcsec$ \citep[][]{Mason2001}, which translates into a physical separation $> 2900$ AU. The proper motions of the two stars are comparable \citep{Hog2000}. Thus, in a scenario where the stars have highly eccentric orbits, a gravitational interaction between them cannot be ruled out. Spatially resolved imaging in the near-IR \citep{Vicente2011} ruled out the presence of stars more massive than $0.22 \ {\rm M_{\odot}}$ at 0.1$\arcsec$, and of any brown dwarf at separations larger than 0.27$\arcsec$. Table \ref{Properties} shows the stellar parameters {of SAO 206462}.

The properties of the disk around SAO 206462 have been constrained via imaging \citep[e.g.][]{Grady2009, Muto2012}, spectroscopy \citep[e.g.][]{Dent2005, Brown2007}, and interferometry \citep[e.g.][]{Fedele2008, Andrews2011}. A large inner dust cavity ($R < 40-50$ AU) has been predicted by SED fitting \citep{Brown2007} and successively imaged at sub- and millimeter wavelengths \citep{Brown2009, Lyo2011}. \citet{Grady2009} and \citet{Muto2012} both resolved the disk in scattered light but no evidence of the inner hole was found down to $\sim$ 42 AU and $\sim$ 28 AU respectively. In particular, the Subaru/HiCiao observations by \citet{Muto2012} revealed two small-scale spiral structures, with the brightest portion of the spirals roughly coinciding with the thermal emission peak at 12 $\mu$m \citep{Marinas2011}. An enhanced emission to SE side and a deficit to north are observed in all sub- and millimeter images and in the 1.1 $\mu$m HST/NICMOS data \citep{Grady2009}. The presence of gas in the inner cavity has been inferred by \citet{Pontoppidan2008} by means of CO rovibrational lines, and \citet{vanderPlas2008} by means of [OI] 6300 $\AA$ line. The presence of a narrow (sub-AU scale) inner dust disk has been inferred by SED modeling \citep{Brown2007, Grady2009} and interferometric N-band observations \citep{Fedele2008}. A variable mass accretion rate of a few $10^{-8} \ {\rm M_{\odot}/yr}$ has been derived by \citet{Sitko2012} from emission lines. The disk inclination is consistently estimated to be $11-14^\circ$ by \citet{Dent2005}, \citet{Pontoppidan2008}, \citet{Andrews2011}, and \citet{Lyo2011}. Periodic variations of the photospheric emission indicates a stellar inclination of $\sim 11^\circ$ \citep{Mueller2011}, suggesting aligned star$-$disk angular momentum. However, \citet{Fedele2008} estimated the inclination of the inner dust disk to be $\sim 60^\circ$. Ellipse fitting of mid-IR imaging suggests a larger inclination also for the outer disk \citep[$i = 40-50^\circ$,][]{Doucet2006, Marinas2011}.     

The paper is organized as follows. Firstly, in Sect.\,\ref{Observations} we describe the observations and the data reduction. Secondly, in Sect.\,\ref{Results} we present the results from the PDI images and, finally, in Sect.\,\ref{Discussion} we discuss a possible scenario for the origin of the structures in the dusty disk of the source.

\begin{table}
      \caption[]{Basic properties of SAO 206462.}
         \label{Properties}
     $$ 
         \begin{tabular}{cc}
            \hline
            \hline
            \noalign{\smallskip}
            Parameter & Value \\
            \noalign{\smallskip}
            \hline
            Right ascension (J2000) & $15^{\rm{h}}$ $15^{\rm{m}}$ $48^{\rm{s}}$.44 $^{\mathrm{\ a}}$   \\
            Declination (J2000) & -37$^{\circ}$ 09$'$ 16$\arcsec$.03 $^{\mathrm{\ a}}$   \\
            $J$ [mag] & 7.279 $\pm$ 0.026 $^{\mathrm{\ b}}$ \\
            $H$ [mag] & 6.587 $\pm$ 0.031 $^{\mathrm{\ b}}$ \\
            $K_{\rm s}$ [mag] & 5.843 $\pm$ 0.020 $^{\mathrm{\ b}}$ \\
            Visual extinction $A_V$ [mag] & 0.3 $^{\mathrm{\ c}}$   \\
            Spectral type & F4Ve $^{\mathrm{\ d}}$ \\
            Luminosity & 7.8 L$_{\odot}$ $^{\mathrm{\ c}}$  \\
            Mass & 1.7 M$_{\odot}$ $^{\mathrm{\ e}}$ \\
            Radius &  $1.4\ {\rm R_{\odot}}$ $^{\mathrm{\ e}}$\\
            Temperature & $6810$ K $^{\mathrm{\ e}}$ \\
            Mass accretion rate & $(0.6 - 4.2) \cdot 10^{-8} \ {\rm M_{\odot}/yr}$ $^{\mathrm{\ f}}$ \\
            Distance & $140 \pm 42$ pc $^{\mathrm{\ g}}$ \\
            Age & 8$^{+8} _{-4}$ Myr $^{\mathrm{\ g}}$ \\
            \noalign{\smallskip}
            \hline
            \noalign{\smallskip}
         \end{tabular}
     $$ 
$^{\mathrm{\ a}}$ \citet{Hog1998}; $^{\mathrm{\ b}}$ \citet{Cutri2003}; $^{\mathrm{\ c}}$ \citet{Andrews2011};  $^{\mathrm{\ d}}$ \citet{Dunkin1997}; $^{\mathrm{\ e}}$ \citet{Mueller2011}; $^{\mathrm{\ f}}$ \citet{Sitko2012}; $^{\mathrm{\ g}}$ \citet{vanBoekel2005}.

\end{table}


\section{Observations and Data Reduction} \label{Observations}
The observations were performed on 2012 July 24 with the high-resolution near-IR NAOS/CONICA \citep[NACO,][]{Lenzen2003, Rousset2003} instrument, mounted on UT4 at the Very Large Telescope (VLT). We used the SL27 camera (pixel scale = 27 mas pixel$^{-1}$) in $HighDynamic$ mode and readout in $Double \ RdRstRd$ mode. SAO 206462 was observed in the {context of a small survey of Herbig Ae/Be disks performed with VLT/NACO in PDI mode on three consecutive nights. The sample contains, among others, HD169142 \citep{Quanz2013} and HD142527 (Avenhaus et al.\ submitted). We obtained images of SAO 206462 in} $H, NB \ 1.64, K_{\rm s}$, and $NB \ 2.17$ filters. The use of narrow band filters is due to the complementary need for exposures with the central star unsaturated, which is unachievable with broad band filters. The total integration times were 3240 and $\sim 3232.9$ s in $H$ and $K_{\rm s}$ filters respectively and 270 s in each narrow band filter. No photometric standard was observed. Instead, we use the central star itself for photometric calibration. Observing conditions were photometric, with {an excellent seeing (varying from 0.58$\arcsec$ to 1.01$\arcsec$), a good coherence time (from 28 to 46 ms), and an average airmass of 1.1}. {The angular resolution of the final images in $H$ and $K_{\rm s}$ band ($\simeq 0.09\arcsec$) is determined from the FWHM of the stellar unsaturated profile in the respective narrow band filter.} Observing settings and conditions are summarized in Table \ref{Settings}. 

\begin{table*}
      \caption[]{Summary of observations. Columns are: filter name, detector integration time (DIT), number of integrations (NDIT) multiplied by integrations per dither position (NINT), number of dither positions, and average airmass, optical seeing, and coherence time. Note that DIT $\times$ NDIT $\times$ NINT $\times$ Dither Position gives the total integration time per position angle of the HWP.}
         \label{Settings}
     $$ 
         \begin{tabular}{ccccccc}
            \hline
            \hline
            \noalign{\smallskip}
            Filter & DIT (s) & NDIT $\times$ NINT & Dither pos. & <Airmass> & <Seeing> & <$\tau_0$> (ms) \\
            \hline
            \noalign{\smallskip}
            $NB \ 1.64$ & 0.5 & 15 $\times$ 3 & 3 & 1.04 & 0.78$\arcsec$ & 35 \\
            $H$ & 0.5 & 90 $\times$ 6 & 3 & 1.03 & 0.78$\arcsec$ & 37 \\
	   $NB \ 2.17$ & 0.5 & 15 $\times$ 3 & 3 & 1.08 & 0.75$\arcsec$ & 38 \\
	   $K_{\rm s}$ & 0.3454 & 130 $\times$ 6 & 3 & 1.12 & 0.70$\arcsec$ & 39 \\
	   \noalign{\smallskip}
            \hline
            \hline
            \noalign{\smallskip}
         \end{tabular}
     $$ 

   \end{table*}

The basic principle of the PDI observing mode is the simultaneous imaging of the linear polarization of the source along two orthogonal directions. With NACO, this is provided by a Wollaston prism which splits the incoming light into ordinary and extraordinary beams, offset by $3.5\arcsec$ along the vertical direction of the detector. A polarimetric mask prevents the overlap of the two beams by alternating open and opaque stripes. A rotatable half-wave plate (HWP) allows an observing strategy with beam switch for the correction of pixel-to-pixel efficiency variations. The polarization measurements at half-wave position angle of 0.0$^\circ$ and -45$^\circ$ yield the Stokes $Q$ parameter, those at angle of -22.5$^\circ$ and -67.5$^\circ$ the $U$ parameter.

The detailed data reduction procedure and correction for instrumental polarization are described by Avenhaus et al.\ (submitted). Shortly, we correct the exposures for dark current and flat field and then perform a row-by-row subtraction of each exposure because of a non-static read-out noise pattern along the rows. Bad pixels are substituted by the mean of the surrounding pixels. The center of the stellar profile in ordinary and extraordinary beams is determined at sub-pixel accuracy by fitting a two-dimensional gaussian to the point spread function (PSF) halo. {For this procedure, values above the linearity regime of the sensor are dismissed}. After that, both beams are extracted and all images are upscaled by a factor 3 and aligned on top of each other.

NACO suffers from instrumental polarization and crosstalk effects \citep[see][]{Witzel2010, Quanz2011}. To correct for the former, we assume the stellar PSF to be largely unpolarized{, and} we impose the count ratio in ordinary / extraordinary beams to be unity. The latter is corrected by scaling the damped signal in Stokes $U$ by a factor defined from the data (see Avenhaus et al.\ submitted). 

The final products of the data reduction are the {tangential and radial Stokes parameters $P_\perp$ and $P_\parallel$}, computed as:
\begin{equation}
\begin{split}
P_\perp = +Q \cos(2\phi) + U \sin(2\phi) \\
P_\parallel = -Q \sin(2\phi) + U \cos(2\phi)
\end{split}
\end{equation}
with
\begin{equation}
\phi = \arctan\left({\frac{x-x_0}{y-y_0}}\right)+\theta
\end{equation}
where $Q$ and $U$ are the Stokes parameters with respect to the sky coordinates \citep[N over E, see][]{Schmid2006}, $\phi$ the polar angle of a given position $(x,y)$ of the detector with respect to the position of the central star, denoted as $(x_0,y_0)$, and $\theta =-3.7^{\circ}$ an offset due to polarization cross-talk effects and a possible misalignment of the HWP rotation within the NACO instrument. By construction, $P_\perp$ and $P_\parallel$ describe the polarization in the {tangential} direction and $\pm 45^{\circ}$ with respect to that, respectively. In systems with {tangential} polarization only, $P_\perp$ is equivalent to $P=\sqrt{Q^2+U^2}$, which denotes the polarized flux, while $P_\parallel$ is not expected to contain a signal and can be used to estimate the error on $P_\perp$. We use this technique because, unlike the $P$ parameter, their computation does not involve the squares of $Q$ and $U$ parameters.

The photometric calibration of the images is performed with the approach described by \citet{Quanz2011}. We estimate the total count rates of the source in $H$ and $K_{\rm s}$ bands (for which the images show a saturated stellar profile) by means of the average of all images in $NB \ 1.64$ and $NB \ 2.17$ filters respectively, taking the different exposure time of the observations and transmission curve of the filters into account. Then, we convert the count rate in each pixel to surface brightness using the 2MASS magnitudes in $H$ and $K_{\rm s}$ bands (see Table \ref{Properties}). This approach intrinsically assumes that the source has the same magnitude in $H$ and $K_{\rm s}$ bands as in their respective narrow bands. However, the presence of a strong Br$\gamma$ line from the star \citep{Sitko2012} may introduce a bias in the calibration of the $K_{\rm s}$ band. Furthermore, the substantial near-IR variability of the source \citep[30\% in the $K$ band,][]{Sitko2012} is not taken into account. We estimate that this technique provides an absolute flux calibration only good to 50\%. 

 \begin{figure*}
   \includegraphics[width=6cm]{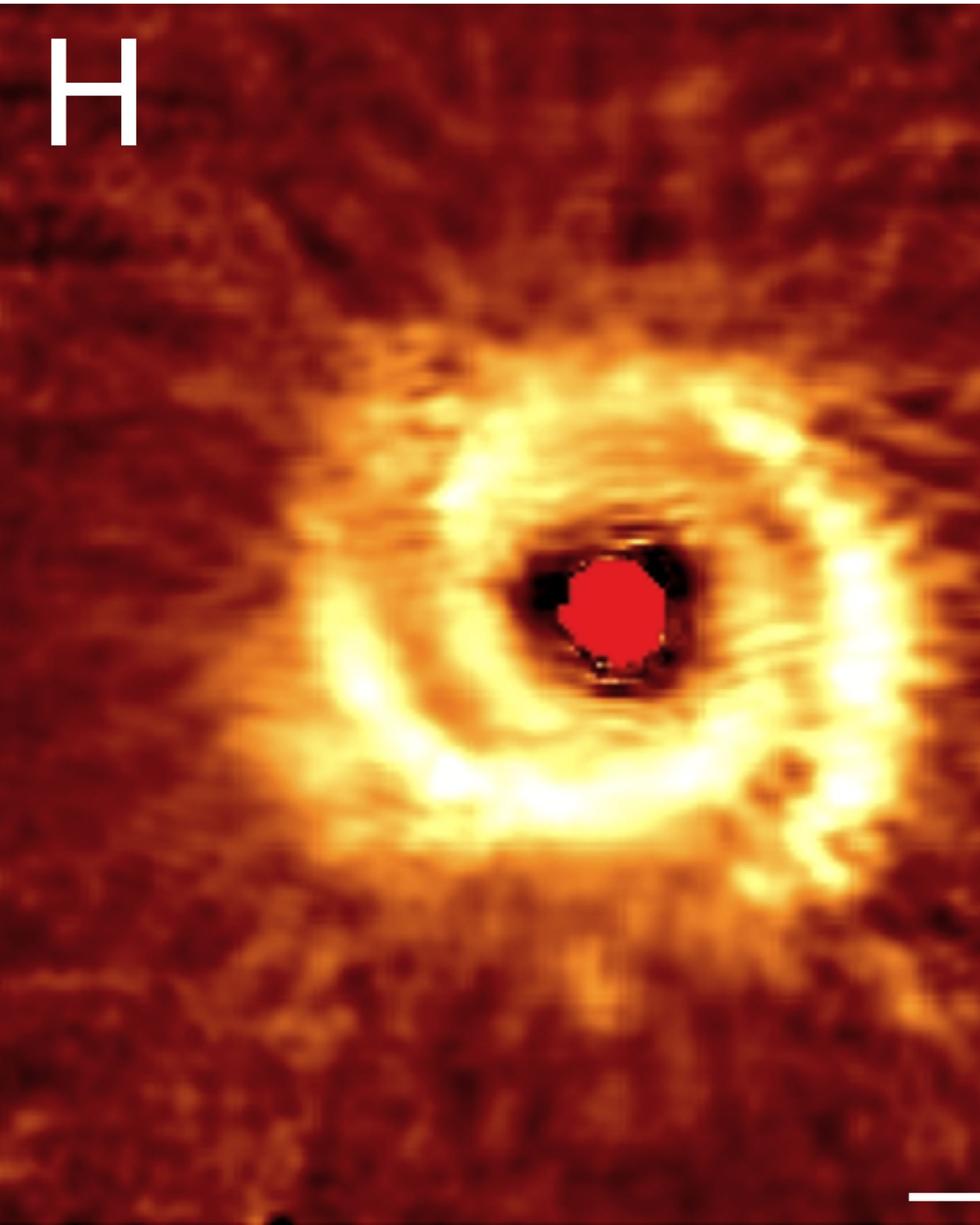}
   \includegraphics[width=6cm]{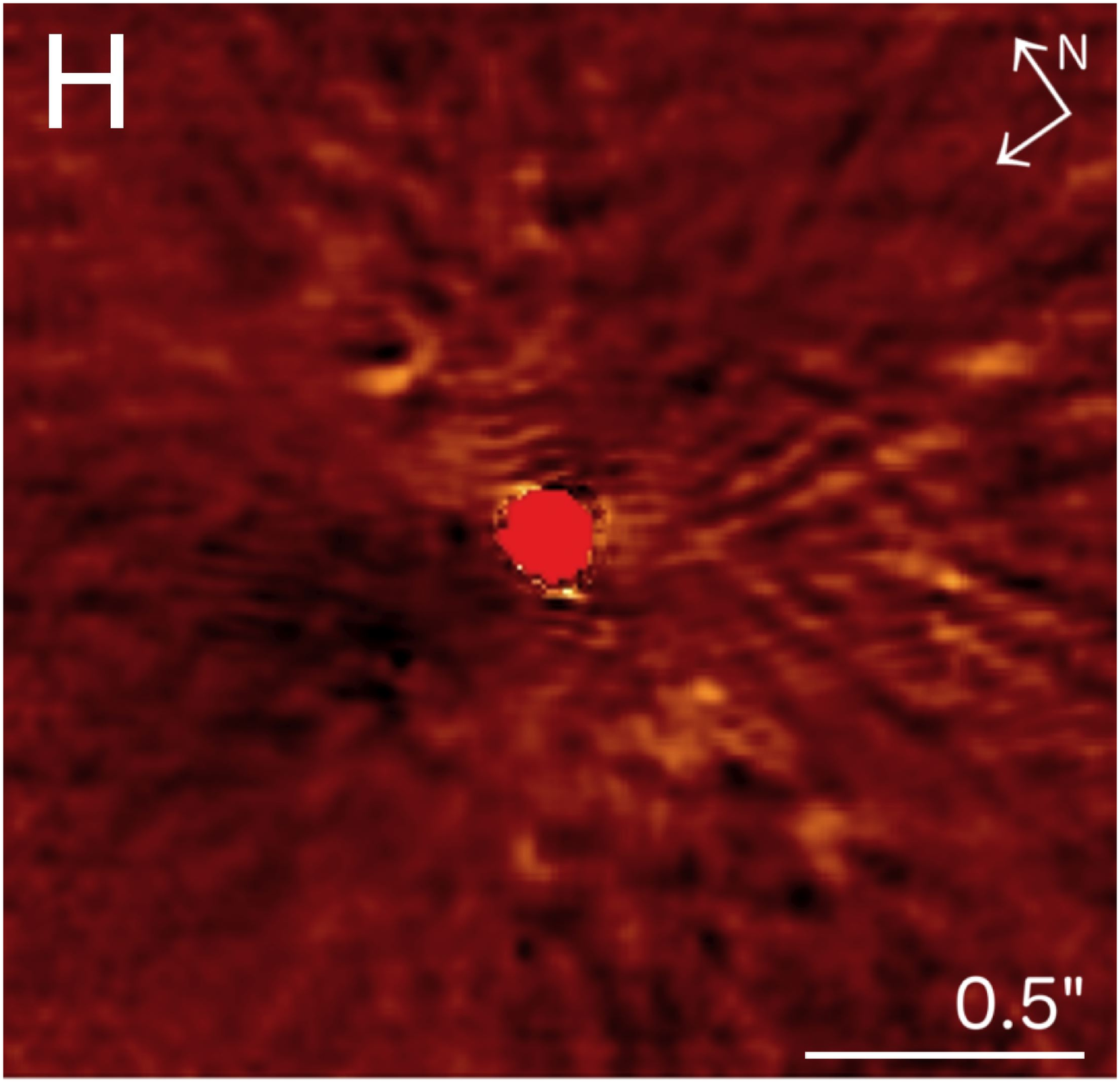}
   \includegraphics[width=6cm]{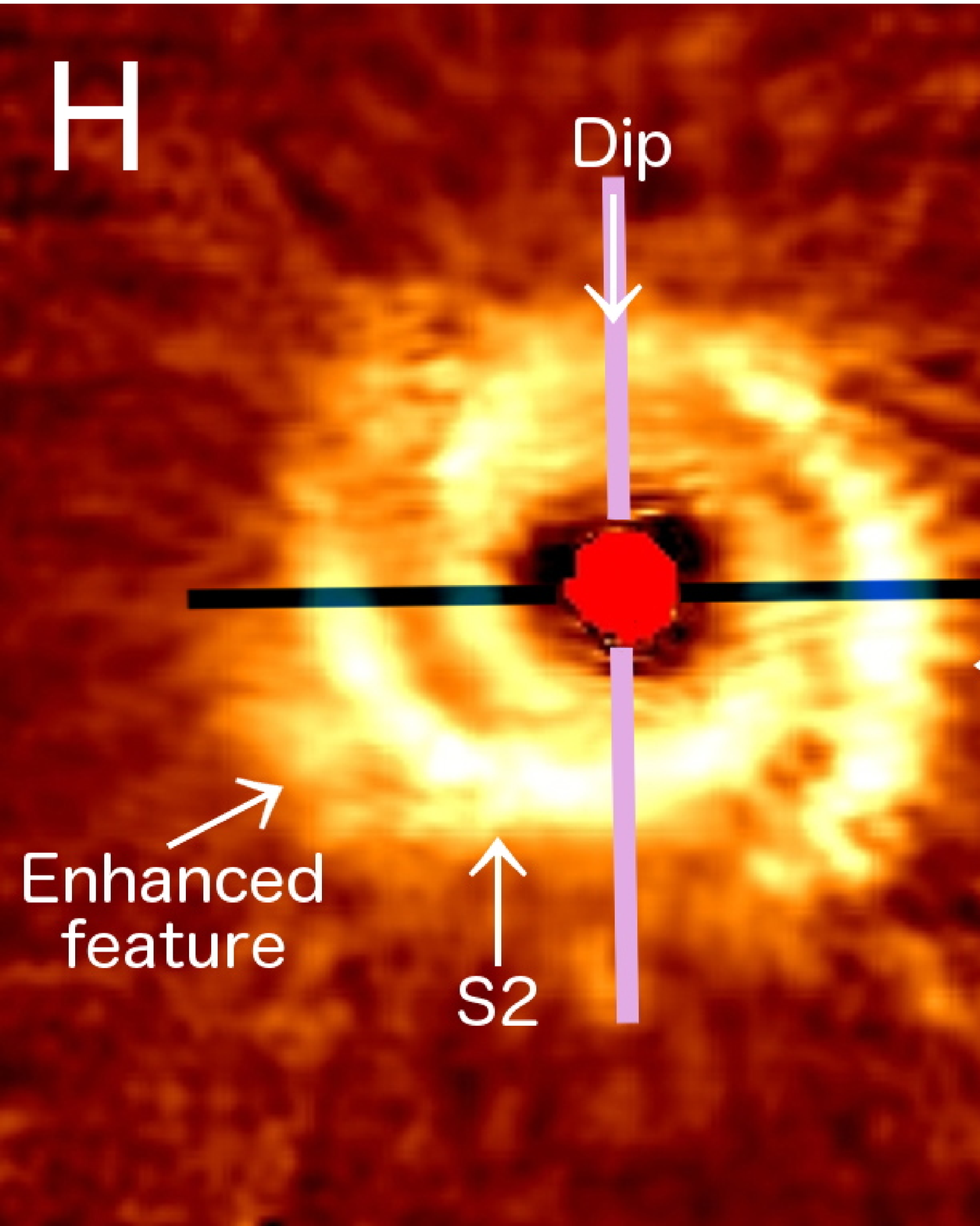}
   \includegraphics[width=6cm]{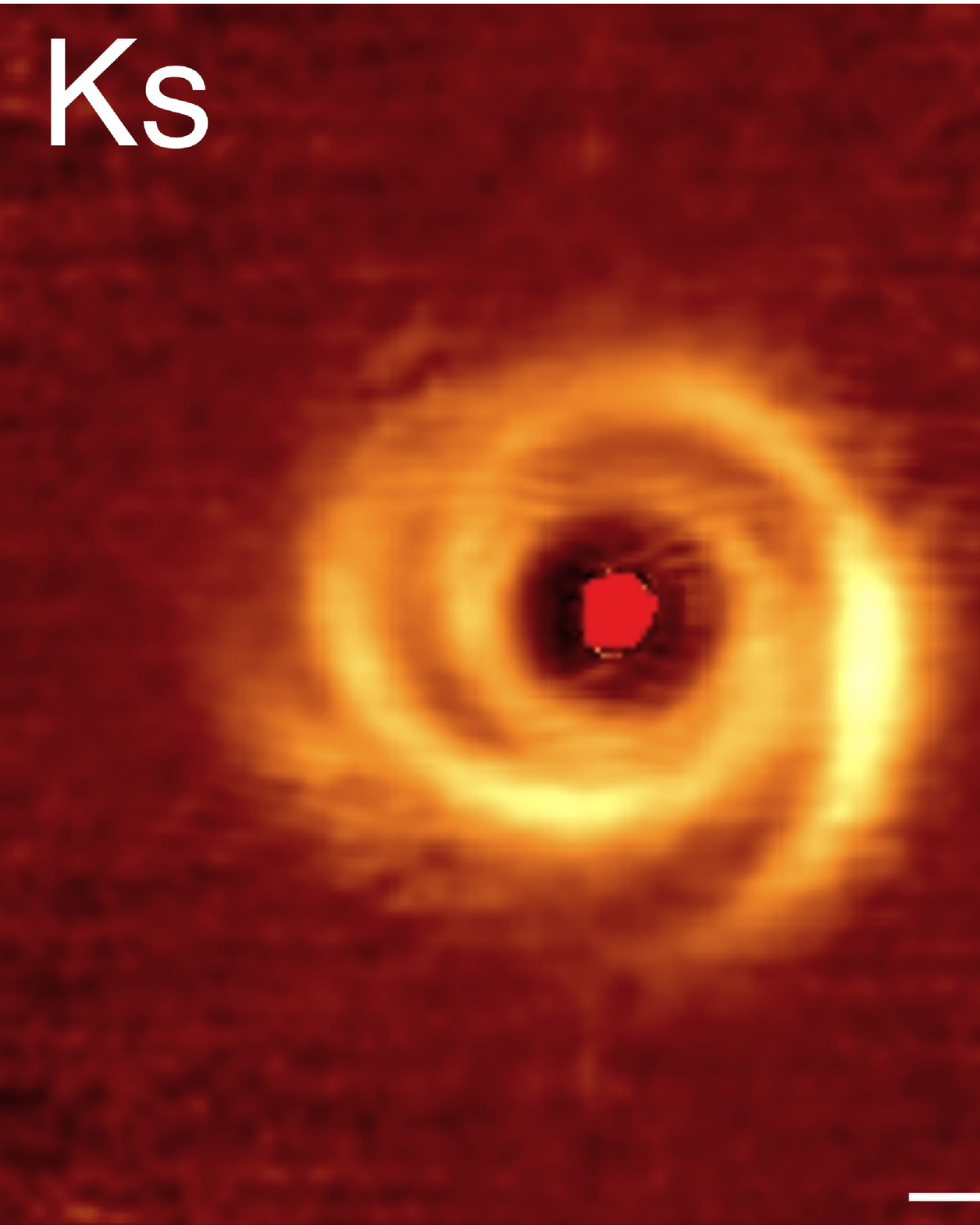}
   \includegraphics[width=6cm]{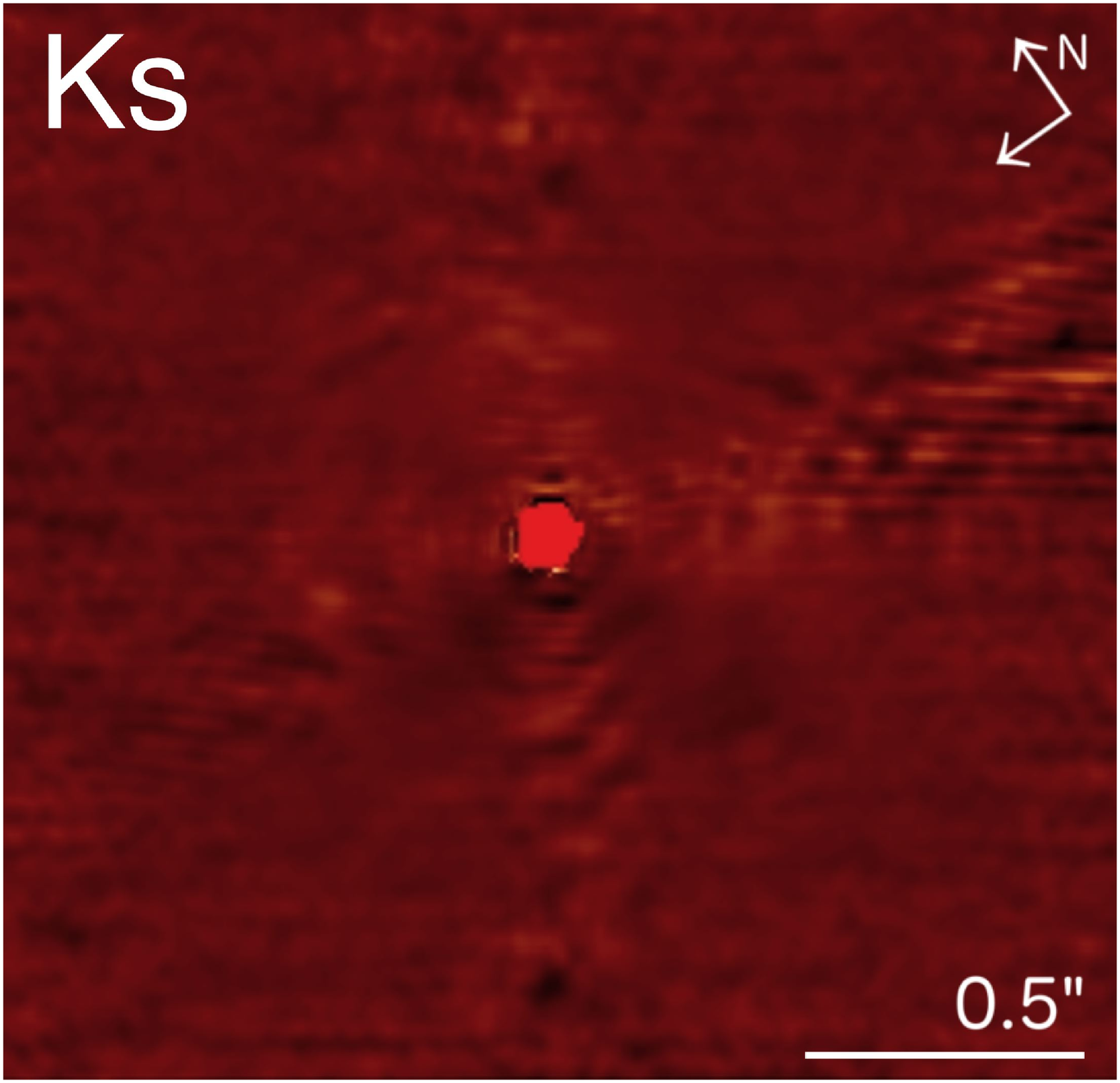}
   \includegraphics[width=6cm]{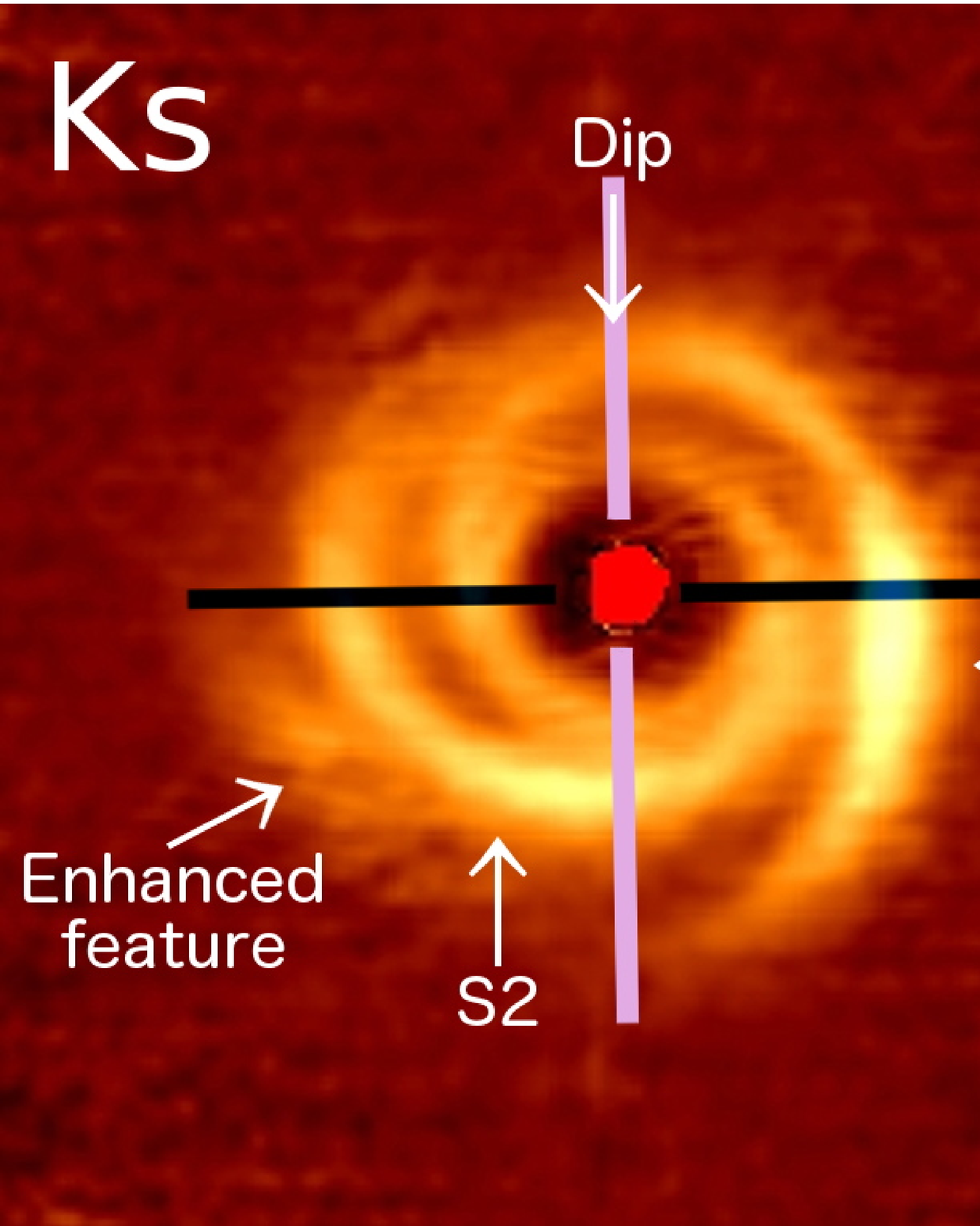}
   \includegraphics[width=6.5cm]{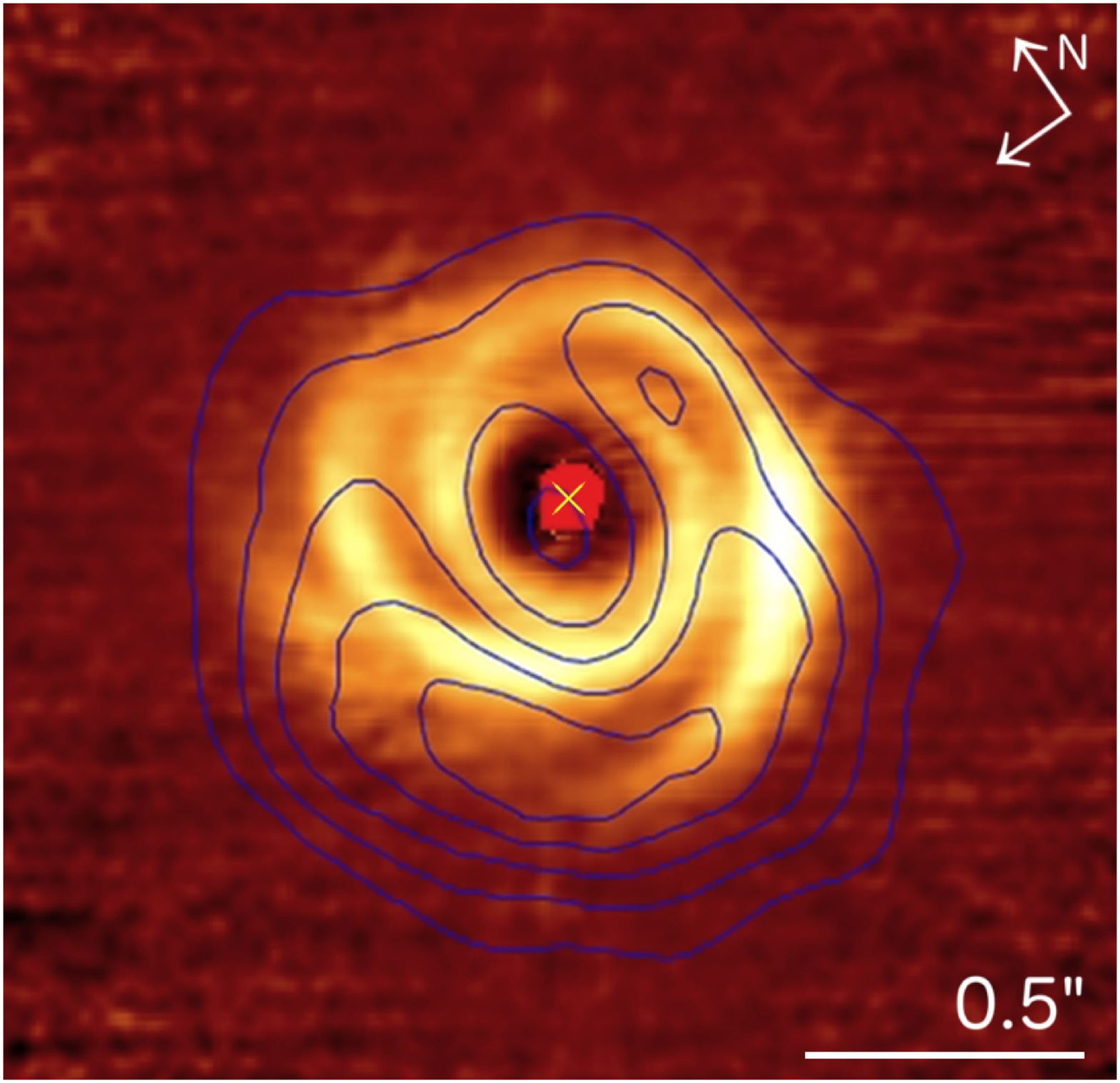}
   \includegraphics[width=6cm]{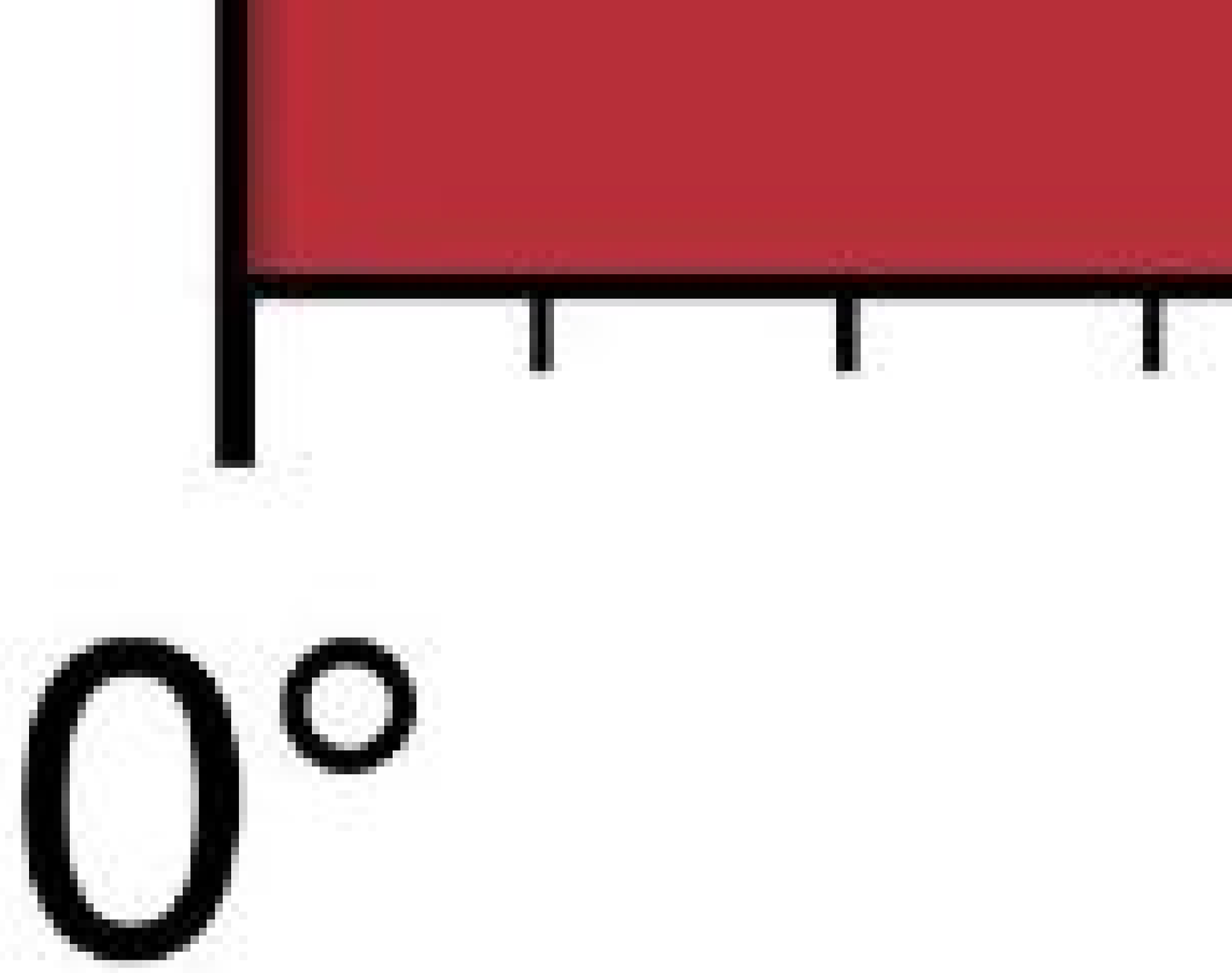} 
   \includegraphics[width=6cm]{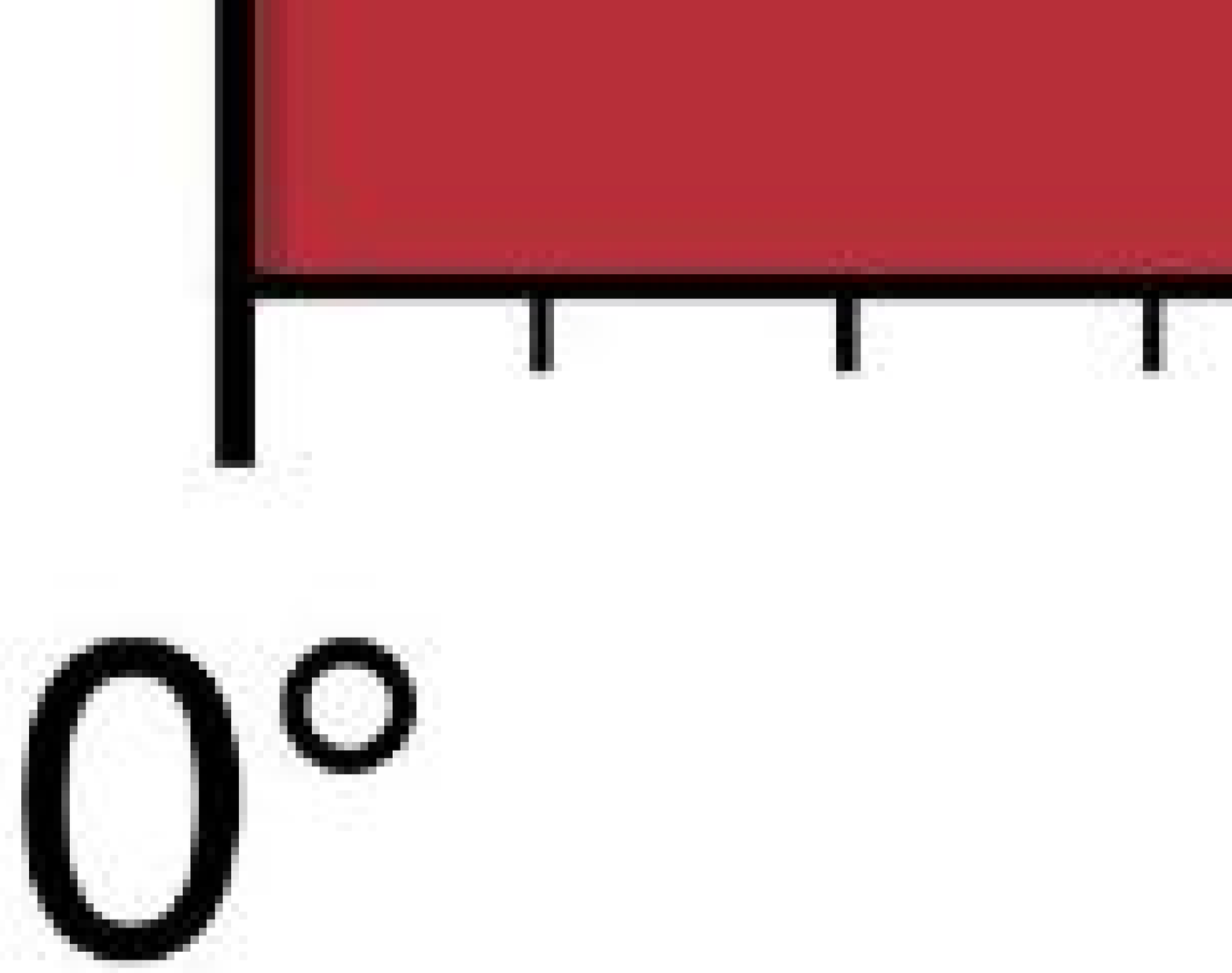}
         \caption{Reflection nebulosity around SAO 206462. First two rows, from left to right: $P_\perp$ image, $P_\parallel$ image, and $P_\perp$ image with {blue} and pink stripes indicating the position of the major and minor axis, respectively. The first row is $H$ band, the second is $K_{\rm s}$ band. Third row, from left to right: comparison between the $P_\perp$ image in $K_{\rm s}$ band and continuum emission from sub-mm observations by \citet{Brown2009}, polar mapping of $H$ and $K_{\rm s}$ band with angles measured with respect to north. Images are {upscaled by a factor 3 to minimize smoothing effects throughout the sub-pixel data shifting and are scaled with $r^2$ to compensate for stellar light dilution. The color scale is linear and arbitrary.} The red central region indicates the area on the detector with non-linear pixels. Continuum contours are drawn at $3\sigma$ intervals starting from $3\sigma$. 
              }
         \label{Images}
   \end{figure*}

\begin{figure*}
   \centering
   \includegraphics[width=14cm]{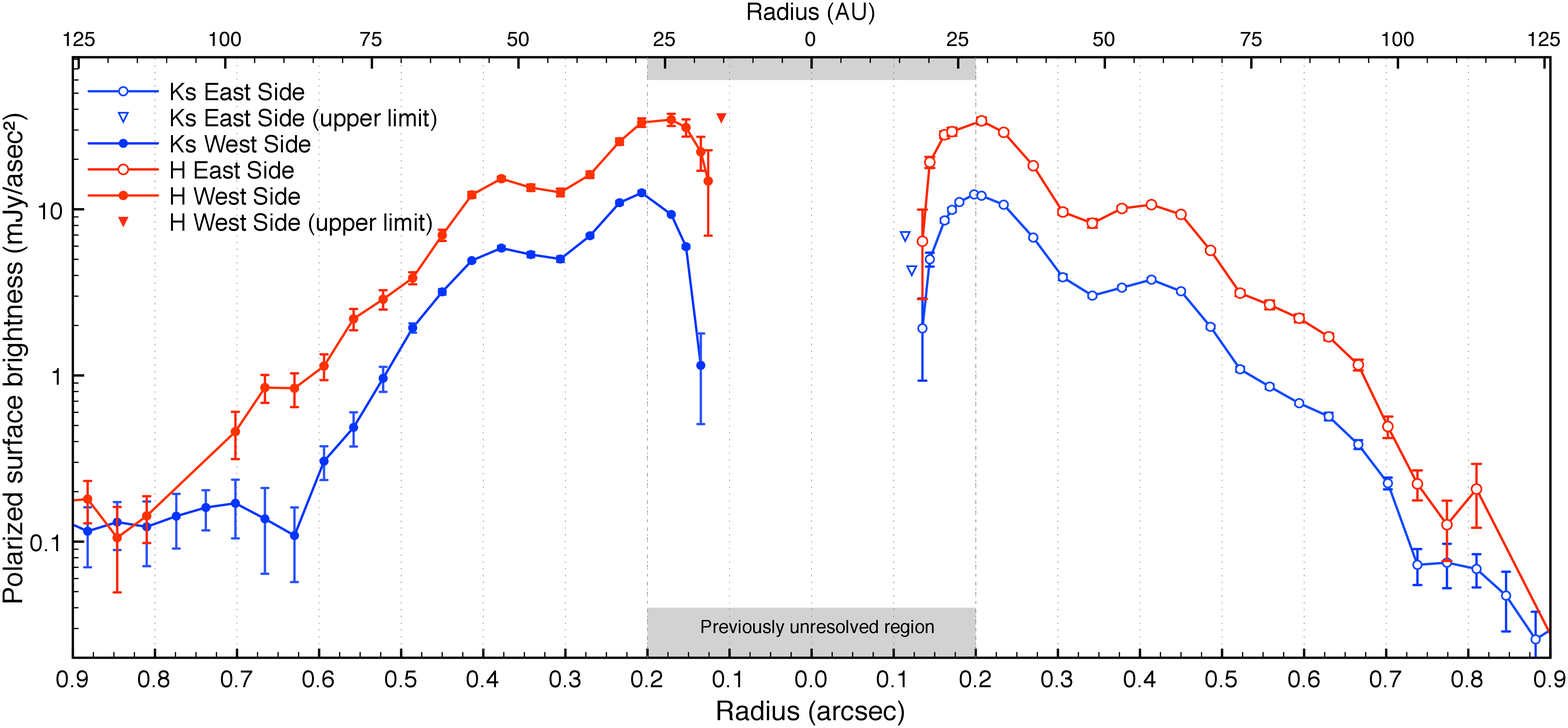}
   \includegraphics[width=14cm]{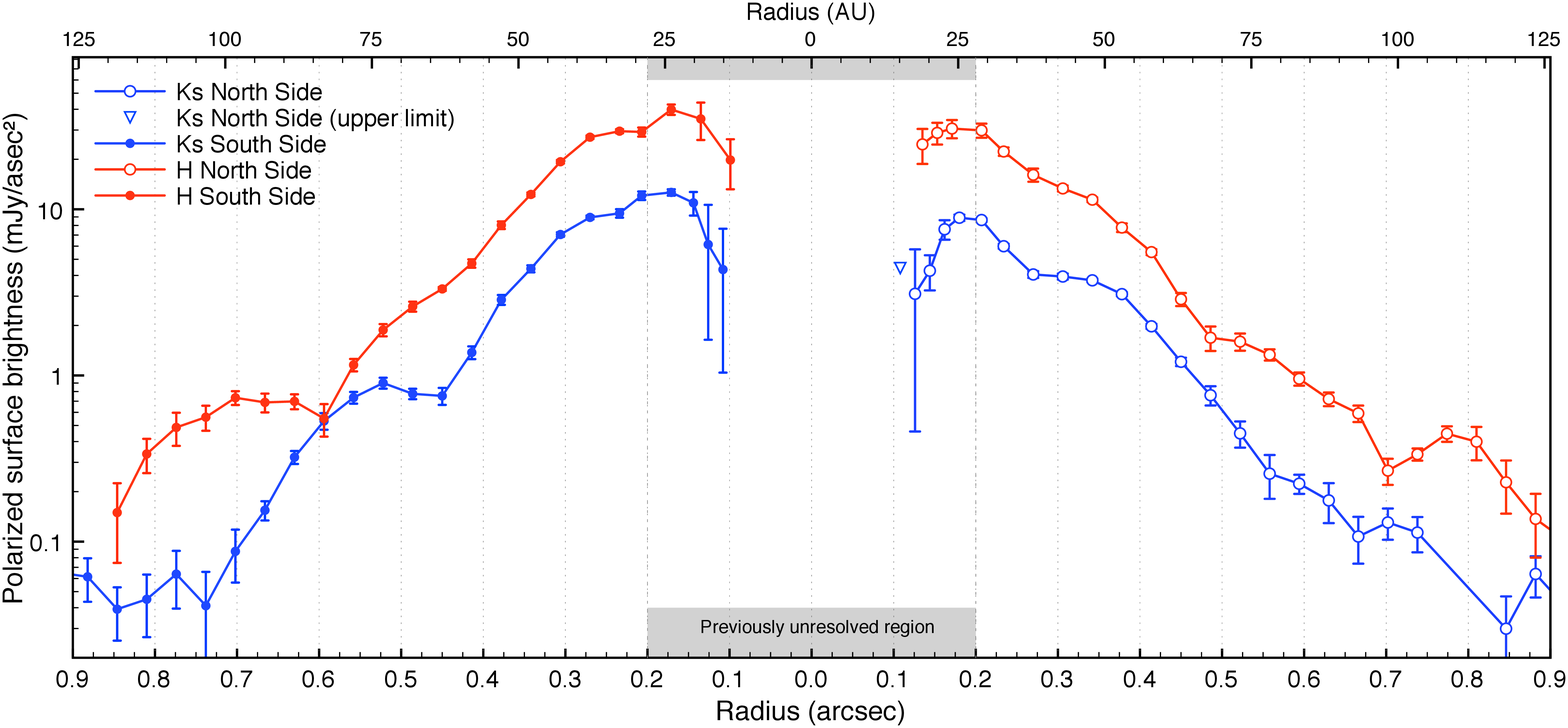}
            \caption{Radial profile of the polarized surface brightness. Top: along the major axis in NE (right half) and in SW (left half) direction for $H$ (red) and $K_{\rm s}$ band (blue). Bottom: along the minor axis in NW (right half) and in SE (left half) direction, namely the far and near disk side, respectively. The grey box indicates the region that our data resolved for the first time. In all profiles (except the $H$ band NW minor axis), the depletion inside this region is evident. Errors are $3\sigma$ noise calculated from the $P_\parallel$ images (see Sect.\,\ref{Observations}) and do not include systematic uncertainties in the photometric calibration.
              }
         \label{Radial}
   \end{figure*}

\section{Results} \label{Results}

Figure\,\ref{Images} shows the final $P_\perp$ and $P_\parallel$ images in $H$ and $K_{\rm s}$ band. {A nebulosity} is clearly detected in both $P_\perp$ images but not in the $P_\parallel$ images. According to the construction of these parameters (see Sect.\,\ref{Observations}), this is indicative of light which is scattered in a {tangentially}-symmetric way. 

The images in $K_{\rm s}$ band are higher-quality because of a better AO correction. Both $P_\parallel$ images reveal partially uncorrected diffraction spikes and read-out noise pattern (see Sect.\,\ref{Observations}). These artifacts are also present in the $P_\perp$ images.

The $P_\perp$ images reveal {the nebulosity} from $\sim 0.1\arcsec$ to $\sim 0.9\arcsec$ (${\sim 14 - 130}$ AU) in both $H$ and $K_{\rm s}$ band. They show three main peculiarities: ($i$) an inner region (inside $0.15\arcsec-0.18\arcsec$) with scattered light depleted down to a few percent of a theoretical continuous distribution (the {\it cavity}); ($ii$) a bright quasi-circular rim at $\sim 0.2\arcsec$ (the {\it ring}); ($iii$) two elongated non-axisymmetric spiral arms extending from the ring to $\sim 0.5\arcsec$ and covering an azimuthal angle of $\sim$ 180$^{\circ}$ each ({\it S1} being that in the W and {\it S2} that in the E, {following} Muto et al.\ 2012). The total intensity of polarized light integrated between $0.1\arcsec$ and $0.9\arcsec$ is 11.7 mJy $\pm \ 50\%$ in $H$ band and 4.3 mJy $\pm \ 50\%$ in $K_{\rm s}$ band. 

In the following, we analyze the polarized brightness distribution of the images, starting from the disk outside the ring and then moving {to the inner $0.2\arcsec$, namely the region that our images resolved in scattered light for the first time.}

\subsection{The brightness distribution: $r > 0.2\arcsec$} \label{Results_all}
The radial polarized brightness is complex and azimuth-dependent because of the presence of spiral arms. Figure\,\ref{Radial} shows the radial profile obtained with 3-pixel wide cuts along the major axis \citep[PA=56$^{\circ}$,][]{Lyo2011} and the minor axis (see orange and pink stripes in Fig.\,\ref{Images}). The relative errors are obtained from the noise estimated from the $P_\parallel$ images (see Sect.\,\ref{Observations}) but do not consider any systematic uncertainty from the absolute flux calibration. No significant difference in the brightness profile is found between $H$ and $K_{\rm s}$ band (apart from the polarized flux being a factor $2.5 - 3$ higher in the former than that in the latter). 

We average the radial profiles over all angular directions neglecting any geometrical effect due to the disk inclination. Since the disk is known to be almost face-on \citep[$11^{\circ}$,][]{Andrews2011}, we assume that this approach does not introduce large systematic errors. The azimuthally-averaged profile of both bands is fitted by a power-law with $\beta = -2.9 \pm 0.1$. However, we find that a spatially separate fit with a broken power-law provides a better match. A slope of $-1.9 \pm 0.1$ is found for the range $0.2\arcsec - 0.4\arcsec$ ($\sim 28 - 56$ AU) and a slope of  $-5.7 \pm 0.1$ ($H$ band) and $-6.3 \pm 0.1$ ($K_{\rm s}$ band) for the range $0.4\arcsec - 0.8\arcsec$ ($\sim 56 - 114$ AU) (see Fig.\,\ref{Radial_average}).

The two spirals are starting from axisymmetric locations on the rim. The contrast of the spiral with the surrounding disk varies from 1.5 to 3.0. S1 appears as the most prominent arm. It covers an angle of $\sim 240^{\circ}$ and shows an enhancement to the SW (almost twice as luminous in surface brightness as the contiguous part of the arm). S2 covers a smaller angle ($\sim 160^{\circ}$) but also shows a "knot" (to the SE, factor 1.5 brighter in surface brightness than the contiguous part of the arm).

From our images, we resolve a narrow dip in the emission along the radial direction at a position angle of 340$^{\circ}$. This feature is very clear on the ring but can be traced out to $0.5\arcsec$ in both bands (see polar mapping of Fig.\,\ref{Images}). A slight deficit can be observed on a larger scale ($\sim 90^{\circ}$) in the NW. A diffuse enhanced emission is also visible in Fig.\,\ref{Images} to the east side. This feature can be traced in both bands from the outer edge of S2 to $\sim 0.7\arcsec$, covering an angle of $\sim 30^{\circ}$. 

The inclination {of the nebulosity} is not strongly constrained by our observations. However, by measuring the {nebulosity} size along the major and the minor axis and assuming overall circular shape, we can rule out values of inclination higher than $25^\circ$.

\subsection{The brightness distribution: $r < 0.2\arcsec$}   \label{Results_cavity}
\begin{figure}
   \centering
   \includegraphics[width=9cm]{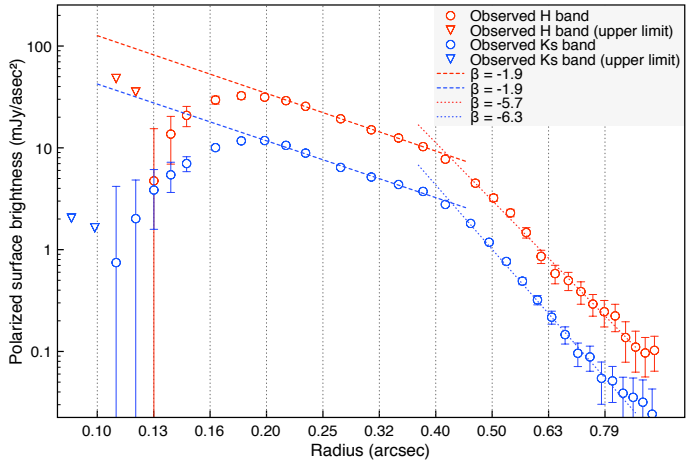}
            \caption{Azimuthally-averaged radial profiles in $H$ (red) and $K_{\rm s}$ (blue) bands and power-law best fits for $0.2\arcsec < r < 0.4\arcsec$ (dashed lines) and for $0.4\arcsec < r < 0.8\arcsec$ (dotted lines). Errors are the same as in Fig.\,\ref{Radial}.         
                 }
         \label{Radial_average}
   \end{figure}

As shown in Fig.\,\ref{Radial}, the polarized flux decreases inward of $0.15\arcsec - 0.18\arcsec$ in both bands.  At $0.12\arcsec$ from the star, the surface brightness is $6\%$ of a continuous flux distribution (as extrapolated from the power-law fit) in $H$ band and $14\%$ in $K_{\rm s}$ band, decreasing to $2\%$ in the innermost region traced by our observations ($0.10\arcsec$). These depletion factors are by far {larger} than the $3\sigma$ error, as shown in the azimuthally-averaged radial profile of Fig.\,\ref{Radial_average}, and therefore significant. 

At the outer edge of the cavity, a quasi-circular ring-like structure is present in both bands. The width of the ring (${\sim 0.09\arcsec}$) is comparable with the angular resolution of the observations, suggesting that the structure is not resolved. The radial location of the brightest part of the ring varies with the azimuthal angle from $0.19\arcsec$ to $0.21\arcsec$. This amount of ellipticity is what is expected from the geometrical projection on the sky (considering $i \sim 10^{\circ} - 20^{\circ}$). Therefore, we do not infer any intrinsic eccentricity for the ring. 

\begin{figure}
   \centering
   \includegraphics[width=9cm]{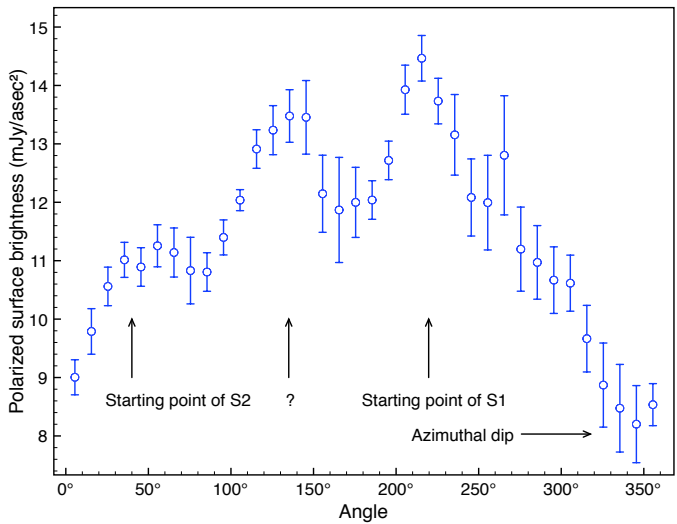}
               \caption{Azimuthal profile of the polarized surface brightness along the ring in $K_{\rm s}$ band, averaged over the $0.17\arcsec - 0.21\arcsec$ radii. Two main bumps are visible. One is associated to the starting point of S1. Errors are the same as in Fig.\,\ref{Radial}, averaged over $0.17\arcsec - 0.21\arcsec$ radii.
                            }
         \label{Azimuthal}
   \end{figure}

In Fig.\,\ref{Azimuthal} we show the azimuthal profile of the ring in $K_{\rm s}$ band, obtained averaging concentric annuli from $0.17\arcsec$ to $0.21\arcsec$. Besides the azimuthal dip ubiquitous along the far minor axis, we identify two features (bumps at PA $\sim 220^{\circ}$ and $\sim 135^{\circ}$). The former is directly associated to S1, since it overlaps with the starting-point of the arm.

\section{Discussion} \label{Discussion}
{Since the $P_\perp$ parameter traces, by construction, the light which is scattered in a {tangentially}-symmetric way, we interpret the reflection nebulosity revealed in the $P_\perp$ images as stellar radiation scattered by the circumstellar disk.} 

In this section, we interpret the results presented in Sect.\,\ref{Results}. In particular, In Sect.\,\ref{Discussion_comparison} we interpret the features observed in the outer part of the disk and compare them with previous works. In Sect.\,\ref{Discussion_cavity}, we focus on the inner cavity and, in Sect.\,\ref{Discussion_implications}, discuss a possible scenario for the origin of the observed features. Finally, in Sect.\,\ref{Discussion_mechanisms} we speculate on the mechanisms responsible for {the} inner cavity and {the} spiral arms and, in Sect.\,\ref{Discussion_planet}, estimate mass and location of the companion {potentially sculpting} the inner disk. 

\subsection{{The outer disk}} \label{Discussion_comparison}
SAO 206462 was detected in scattered light with HST/NICMOS at 1.1 and 1.6 $\mu$m by \citet{Grady2009} and, later, in polarized light with Subaru/HiCIAO in $H$ band by \citet{Muto2012}. The latter were the first to resolve the spiral structure. The outer extent inferred by our observations ($0.9\arcsec$) is in good agreement with their estimate. The azimuthally averaged radial profile fitting our observations (power-law with $\beta = -2.9$) is in the range of typically observed values for Herbig stars, both in scattered light \citep[e.g.][]{Grady2007} and polarized light \citep[e.g.][]{Quanz2011}. The value is in agreement with what was inferred for SAO 206462 by \citet[$r^{-3}$ as a representative average]{Muto2012}. The total polarized light intensity measured in the $H$ band (11.7 mJy) is a factor 1.15 higher than the measurement by \citet{Muto2012} but the results agree within the error bars. Furthermore, a small part ($\sim 10\%$) of the total intensity is from the inner region ($< 0.2\arcsec$) that is masked in the images by \citet{Muto2012}, reinforcing the agreement. Finally, the total polarized light intensity in $K_{\rm s}$ band (4.3 mJy) was never measured before.

The spiral arms are by far the most tantalizing feature of the disk. Spiral arms have been observed in a few other protoplanetary disks. Among them, {we point out} AB Aur \citep{Hashimoto2011}, HD 100546 \citep{Grady2001}, HD142527 \citep[Avenhaus et al.\ submitted]{Casassus2012}, and MWC 758 \citep{Grady2013}. Since only very recent techniques {are capable of resolving} these structures, it is difficult to infer how common spiral arms are among protoplanetary disks. Because PDI observations trace the stellar radiation which is scattered by a thin surface layer of the disk, the spiral arms detected with this technique can either constitute intrinsic changes of the disk grain properties or small local variations of the disk geometry, which modify the incidence angle of the photons. In particular, small ripples in the disk scale height may be sufficient to explain the light contrast shown by the spirals around SAO 206462 with the contiguous disk ({a factor} 1.5 to 3.0). Disks with this configuration might not show spiral structures in their thermal images.  

The brightness deficit in the NW was also observed by \citet{Muto2012}. The same deficit was observed in scattered light by \citet{Grady2009} at 1.1 $\mu$m. Curiously, they did not resolve it at 1.6 $\mu$m, whereas our $H$ band image (of polarized light) does. \citet{Lyo2011} inferred from CO observations that the SW side is receding. Spirals are typically trailing and, thus, as remarked by \citet{Muto2012}, the region of the deficit may represent the far side of the disk. In particular, the dip in polarized light at 340$^{\circ}$ is roughly coincident with the minor axis (326$^{\circ}$). Given this, the narrow dip and the general deficit can be ascribed to depolarization. Since the degree of linear polarization is maximized at $\sim 90^{\circ}$ scattering angle \citep[e.g.][]{Draine2003, Min2012}, polarized images of inclined disks are indeed expected to show minima along the near and far side of the disk \citep[see case of MWC480,][]{Kusakabe2012}. However, no deficit is observed on the near side. This might indicate that dust grains preferentially forward-scatter the radiation, thus compensating for the depolarization effect.

The horseshoe structure revealed from thermal images by \citet{Brown2009} and \citet{Lyo2011} has {a} possible connection with the polarized light (see comparison in Fig.\,\ref{Images}). The bright part of the horseshoe is in the south, where also the luminous part of the disk in scattered light is {located}. Interestingly, the peak of the thermal emission, in the SE, seems to lie in a darker region of the PDI images, namely between the tail of S1 and the enhanced feature in the east. However, the nature of this enhanced feature is unclear. The spread deficit in polarized light, in the north, is also resolved in the thermal images, suggesting that the lack of polarized light can be (partially) ascribed to some geometrical properties of the local dust, such as a diminished scale height.

\subsection{The cavity and the ring} \label{Discussion_cavity}
{The inner region ($r < 0.2\arcsec$) of the disk was never resolved before in scattered light. Therefore, we dedicate the rest of the section to the features identified therein.} We interpret the drop in surface brightness inside $0.15\arcsec - 0.18\arcsec$ as a real depletion of small dust grains. In the following, we refer to the region showing this depletion as the cavity. This definition is meant to indicate a region showing an abrupt discontinuity in the dust radial profile but does not {imply} that this region is completely void of dust. The {ring} revealed at the outer edge of the cavity represents the most luminous region of the disk and we interpret it as the inner wall of the outer disk which is directly illuminated by the central star. Thus, we infer the outer edge of the cavity from the (unresolved) ring radial profile along the major axis. We find that the cavity extends from the instrumental inner working angle ($\sim 14$ AU) to {${28 \pm 6 \ {\rm AU}}$. This cavity edge is in good agreement with what was recently inferred for this source by \citet{Maaskant2013} with Q-band ($24.5 \ \mu$m) imaging (30 AU).

Interestingly, the outer edge of the observed cavity considerably differs from that predicted by SED fitting \citep[$R = 45$ AU,][]{Brown2007} and that imaged at sub-millimeter \citep[$R = 39$ AU and $46$ AU,][]{Brown2009, Andrews2011} and millimeter wavelengths \citep[$R$ = 50 AU,][]{Lyo2011}. {Even though the angular resolution of (sub-)millimeter observations is coarser than that of our PDI images, this discrepancy cannot be explained by observational effects. In fact, downgrading the resolution of our images acts to fill in the gap, thus raising the difference between the datasets.} Discrepancies between thermal and polarized light were already noticed by \citet{Muto2012}. However, they found no evidence of discontinuity {at radii as small as 28 AU}.

The inner (sub-AU scale) $60^{\circ}$-inclined dust belt inferred by \citet{Fedele2008} should cast a shadow on the outer disk wall, presumably on two distinct axisymmetric points of the {ring}. However, we do not infer the presence of shadow cones, neither on the {ring} nor on the external part of the disk. This might not appear as a contradiction, since \citet{Espaillat2010} showed that, for typical pre-transitional disks, significant portions of the cavity wall might be out of the shadow casted by the inner wall.

Azimuthal asymmetry in transitional disks are now intensively studied because of the recent discovery of highly azimuthal-asymmetric disks \citep[e.g.][]{Brown2009, vanderMarel2013}. \citet{Lin2011} have shown that the gas accumulated at the outer edge of a cavity carved out by a planet may become unstable and, thus, generate vortices. In the assumption that small dust grains at the cavity edge are perfectly coupled to the gas, dust accumulation in rings might be detected from PDI observations and could possibly be a signpost of the presence of a planet in the cavity. The bump in the azimuthal profile of Fig.\,\ref{Azimuthal} at PA $\sim 220^{\circ}$ is associated with the starting-point of S1. Curiously, the position of this bump is consistent with that of the azimuthal asymmetry shown by CO lines in the inner 25 AU \citep[at PA $\sim 240^{\circ}$,][]{Pontoppidan2008}. The other bump (at PA $\sim 135^{\circ}$) is not in evident relation with other features of our images and should be considered as possible location for small dust grain concentration in the rim. Finally, the azimuthal dip at PA $\sim 340^{\circ}$ can be explained by the depolarization effect, as described in Sect.\,\ref{Discussion_comparison}.

\subsection{Small and large dust grains at the cavity edge} \label{Discussion_implications}

\begin{figure*}
   \centering
   \includegraphics[width=14cm]{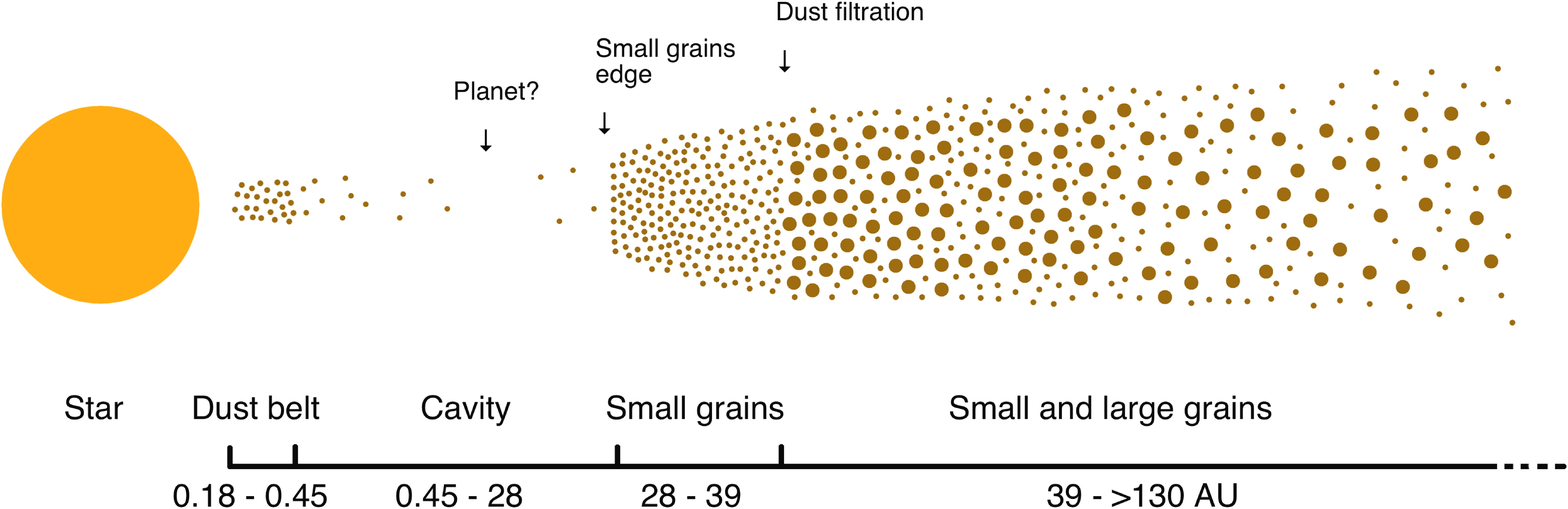}
               \caption{Illustrative sketch showing the dust radial distribution in the disk around SAO 206462, as proposed in this work according to data from this paper, \citet{Fedele2008}, and \citet{Brown2009}. No speculation on the vertical structure is contained in the sketch.
               }
         \label{Sketch}
   \end{figure*}

The size of the cavity of many transitional disks has been inferred from sub-millimeter observations \citep[see e.g.][]{Andrews2011}. However, {the presence of small dust grains well inside millimeter-grains cavities has been inferred by means of PDI observations \citep[e.g.\ UX Tau A; SR21,][]{Tanii2012, Follette2013} and modeling of multi-wavelength observations \citep[e.g.\ HH 30,][]{Madlener2012}}. On the other hand, some objects \citep[e.g.\ GM Aur,][]{Graefe2011} do not show any dissimilarity in the cavity size for different grain sizes. Since {PDI and thermal images} trace different components of the dust ($\mu$m-sized grains in the scattered light images, mm-sized grains in the thermal images), this discrepancy can provide important information on the dust distribution in the inner regions. 

\citet{Rice2006} have shown that, at the outer edge of a disk gap cleared by a planet, a local pressure bump can act as a filter, allowing small ($\lesssim 10 \ \mu$m-sized) particles to drift inward but holding back larger grains. In order for the SED to show the near- to mid-IR dip typical of transitional disks, the region inside the cavity must contain optically thin dust. However, \citet{Zhu2012} argued that dust filtration alone cannot explain the observed spectral deficit (invoking additional processes such as dust growth). \citet{Dong2012} put the three observational evidence together (spectral deficit, cavity in sub-mm images, and continuous radial distribution of polarized light) and proposed a model with dust filtration and flat or inward-decreasing surface density profile inside the cavity (consistent with dust growth).

Nevertheless, our observations replace, for the particular case of SAO 206462, one element: the small dust grains are not continuously distributed down to the sublimation radius but show an abrupt discontinuity. Since the location of the observed discontinuity ($0.15\arcsec$) is comparable with the typical inner working angles of PDI ($0.1\arcsec - 0.2\arcsec$), we argue that the case of SAO 206462 may not be unique. The presence of a depletion for small dust grains represents a direct, qualitative explanation for the spectral deficit, without necessarily invoking simultaneous dust growth inside the cavity. Disk cavities were already detected in PDI in a number of other objects \citep[e.g.\ HD 100546 and HD 169142,][]{Quanz2011, Quanz2013}. Additional PDI observations (with inner working angle of $0.1\arcsec$ or lower) of a sample of transitional disks are needed to speculate whether two families of geometries (abrupt and continuous radial dust profiles) may exist. 

Dust filtration permits the presence of small grains inside the edge observed at millimeter wavelengths but cannot explain why these grains are held back at a different, inner location. A possible explanation invokes the gas distribution inside the cavity. \citet{Pinilla2012} have shown that the outer edge of the gaseous halo carved out by a planet may not coincide with the location of the pressure bump generated by the same planet. They show how the pile-up of the dust occurs, depending on the planet mass, at radii 1.4 to 2.0 times farther from the planet than the gap outer edge in the gas. Only particles larger than a certain minimum size ({typically at sub- to millimeter scale at distances of $20-50$ AU}) are perfectly trapped in the pressure maximum. {We expect that the small ($\mu$m-size) particles at $R \sim 40$ AU are perfectly coupled to the gas and can, thus, be dragged along with it. To show this, we calculate the Stokes number St, defined as:}
\begin{equation}
{\rm St}(a, R)=\frac{\rho_{\rm S}(R) \cdot a}{\Sigma_{\rm g}(R)} \cdot \frac{\pi}{2}
\end{equation} 
{\citep{Birnstiel2010a} where $a$ is the particle size, $\rho_{\rm S}$ the solid density of dust particles, and $\Sigma_{\rm g}$ the gas column density.  We calculate St for $\mu$m-size particles located at $R = 40$ AU using a typical $\rho_{\rm S}=1 \ {\rm g/cm^3}$ and $\Sigma_{\rm g}$ as constrained for this disk by \citet{Andrews2011}:} 
\begin{equation}
\Sigma_{\rm g}(R)=22\,\left(\frac{R}{{\rm 55\,AU}}\right)^{-1} \cdot \exp \left(-\frac{R}{\rm 55\,AU}\right)\ {\rm g/cm^2}
\end{equation}
{We find that St$(a=1\ \mu {\rm m},\ R=40\ {\rm AU})\simeq 10^{-5}$. Since strong interplay with the gas is expected for particles with St $\ll 1$, this result suggests that $\mu$m-size dust grains at the cavity edge of mm-size grains are very well coupled to the gas.}

In a scenario where a planet is carving out the disk (see Sect.\,\ref{Discussion_mechanisms}), the observed inner cavity of $\mu$m-size dust grains at 28 AU might coincide with the outer edge of the gaseous cavity, whereas the mm-size particles are piled-up at the pressure bump located at 39 AU. However, the inner region is not void of gas, as inferred from the presence of CO ro-vibrational lines as far as 25 AU \citep[considering $d=142$ pc,][Carmona et al.\ in prep.]{Pontoppidan2008}. This may not appear as a contradiction of the proposed scenario if we assume that the gas is carved out in a narrow ring (consistently with the presence of a single planet) because CO line profile fitting may not reveal such a fine structure. The observed depletion (one order of magnitude) of CO gas in the central region of the disk \citep{Lyo2011} might also support a scenario where the gas is partially cleared. In fact, the tidal action of a planet is also expected to significantly deplete material in the region inside the location of the planet \citep{Tatulli2011}. Furthermore, the presence of gas and small dust grains inside the cavity is required to explain the inner dust belt and the substantial accretion rate of the source. In Fig.\,\ref{Sketch}, we depict the suggested dust radial distribution around SAO 206462. 
 
\subsection{What can cause cavity and spiral arms?}  \label{Discussion_mechanisms}
Spiral arms and large inner cavities are among the most intriguing features observed in protoplanetary disks. These two features (both shown by the dusty disk around SAO 206462) can provide insight into the {dynamical} processes occurring in the disk.  

{\it Inner cavity}. One possible process responsible for the clearing of a large inner region of disks is the interaction with orbiting companion(s) \citep[e.g.][]{Rice2003}. \citet{Dodson-Robinson2011} have shown with hydrodynamical simulations that a multiple planets system is required to open holes as large as the millimeter observations suggest. However, \citet{Pinilla2012} also considered modeling of dust evolution demonstrating that the gap carved out by a single planet can be much larger than expected by means of pure gas hydrodynamical simulations. Photoevaporation can also generate disk inner holes. \citet{Alexander2007} argued that UV photoevaporation can induce holes only in systems with accretion rates $\lesssim 10^{-10} \ {\rm M_{\odot} \ yr^{-1}}$. Thus, the substantial accretion rate shown by SAO 206462 \citep[a few $10^{-8} \ {\rm M_{\odot}/yr}$,][]{Sitko2012} rules out this possibility. Dust grain growth has also been proposed to explain the observed cavities \citep{Dullemond2005}. However, this process is expected to produce smooth radial profiles, inconsistent with our PDI observations. Furthermore, the growth of $\mu$m-size particles should give rise to a family of mm-size grains, which is not observed at (sub-)mm wavelengths \citep{Brown2009, Lyo2011}. \citet{Birnstiel2012} showed indeed that large cavities cannot be caused by grain growth alone. Finally, the role of magnetorotational instability (MRI) in disk clearing has been discussed by \citet{Chiang2007}. However, \citet{Dominik2011} argued that the dust grains cannot be held back by radiation pressure alone.
In addition, the presence of a thick dust belt close to the star \citep{Fedele2008} rules out any mechanism operating with inside-outside modality (namely, the photoevaporation and the MRI instability).  

{\it Spirals.} Disk$-$planet interaction can result in spiral density waves \citep[e.g.][]{Tanaka2002} and possibly generate both a disk gap and a spiral arm structure \citep{Crida2006}. Alternatively, gravitational instability has been proposed as mechanism of spiral wave excitation in disks \citep[e.g.][]{Durisen2007}. Disks are considered gravitationally unstable at a certain radius $r$ if the Toomre parameter $Q(r) \lesssim 1$. To speculate about the stability of the disk around SAO 206462, we use a global $Q$ as calculated by \citet{Kamp2011}: 
\begin{equation}
Q=\sqrt{\frac{8k}{\mu m_{\rm H}G}}\sqrt{\frac{\sqrt{R_*}T_*M_*}{M^2_{\rm disk}(2-\epsilon)^2}}R^{1/4}_{\rm in}\left(\left(\frac{R_{\rm out}}{R_{\rm in}}\right)^{(2-\epsilon)}-1\right)\left(\frac{R_{\rm out}}{R_{\rm in}}\right)^{(\epsilon-7/4)}
\end{equation}
We use the stellar radius $R_*$, temperature $T_*$, and mass $M_*$ assumed by \citet{Mueller2011}, a disk mass $M_{\rm disk}$ of 0.026 $\rm M_{\odot}$ \citep{Andrews2011}, disk outer radius $R_{\rm out}$ of 220 AU \citep{Lyo2011}, and inner radius $R_{\rm in}$ from this dataset. We vary the column density power-law $\epsilon$ over a range of realistic values \citep[0.5 to 1.5,][]{Kamp2011}. We find that $Q$ always exceeds 25, meaning that the disk is (globally) highly stable. To enter the unstable regime, the disk would have to be more than twenty times more massive. However, the degree of instability necessary to generate small wiggles on the disk surface (see Sect.\,\ref{Discussion_comparison}) may be lower than what is usually assumed as an edge of the two regimes. Finally, formation of spiral arms is expected in disks around binary stars because of strong tidal forces \citep[e.g.][]{Boss2006}. The physical separation between SAO 206462 and its visual companion HD135344A is at least 2900 AU. In the case of highly eccentric orbits ($e=0.9$) and face-on orbital plane, the semi-minor axis of this potential binary system can be as small as $\sim150$ AU. \citet{Duchene2010} showed that binaries wider than 100 AU are indistinguishable from single stars in terms of their circumstellar disks and planetary systems. Nevertheless, we cannot rule out a marginal gravitational interaction between the two stars. This study is beyond the scope of the current work.

It turns out that the interaction with an orbiting companion is the only process capable of explaining both the inner disk clearing and the spiral arms structure. Any other known mechanism cannot account for at least one observational evidence. However, we cannot claim that a potential companion responsible for the inner clearing is also responsible for the spiral excitation. In fact, \citet{Muto2012} and \citet{Grady2013} estimated the location of the possible planet exciting spiral arms (of SAO 206462 and MWC 758, respectively) to be outside of the spirals themselves. \citet{Grady2013} argued that such a planet could not account for the cavity observed around MWC 758.  

In the following section, we focus on the possible companion responsible for clearing. The scenario for which the spiral arms are excited by the same (or by another) companion is not further discussed in this paper.

\subsection{An orbiting companion?} \label{Discussion_planet}
In the intriguing prospect that a single object is responsible for the observed large inner hole, we constrain the location and the mass that such a companion should have. We use the results by \citet{Dodson-Robinson2011} and \citet{Pinilla2012} to determine a testable range of realistic locations for that object and, then, infer its mass. 

The gap width in the disk is regulated by the tidal radius, or Hill radius, of the companion, defined as:
\begin{equation}
R_H=r_p \left( \frac{M_p}{3M_*} \right) ^{1/3}
\end{equation}
being $r_p$ and $M_p$ the planet location and mass and $M_*$ the stellar mass. Dodson-Robinson \& Salyk (2011) showed that tidal interactions with planets are very fast at scattering disk particles within $\sim 3 \ {\rm R_H}$ ($\lesssim 10^4$ yr) while the scattering timescale for particles at ${\sim 5}\ {\rm R_H}$ approaches 1 Myr. These results argue against the single planet carving out the gap, because of the inability of reproducing the hole sizes observed at millimeter wavelengths. However, \citet{Pinilla2012} predicted that the accumulation of large dust grains can occur at radii 1.4 to 2.0 larger than the outer edge of the gaseous gap (i.e.\ $\sim 7$ to $\sim 10\ {\rm R_H}$). 

Inspired by this statement, we use the speculated outer edge of the gas cavity ($R=28$ AU, as inferred from our observations of small dust grains, see Sect. \ref{Discussion_cavity}), and impose it to be the outer edge of a gap with size (i.e.\ distance planet$-$edge) of $\kappa_{\rm gas}=3$ to 5 times ${\rm R_H}$. Moreover, we use the observed cavity dimension for mm-size dust particles \citep[$R=39$ AU, from][]{Brown2009} and impose it to be the outer edge of a gap of $\kappa_{\rm dust}=7$ to 10 $\rm R_H$ \citep[depending, in turn, on the mass planet, see][]{Pinilla2012}. By varying $\kappa_{\rm gas}$ and $\kappa_{\rm dust}$ over a discrete range of values, we find a family of solutions (see Fig.\,\ref{Planet}). A {potential} companion responsible for the observed geometry should be located at a radius spanning from 17 to 20 AU with a mass between 5 and 15 $\rm M_J$, being, thus, a giant planet. These companion masses are not in contradiction with the upper mass limit inferred by \citet{Vicente2011} with VLT/NACO narrow-band imaging ($\sim 230 \ {\rm M_J}$ at 14 AU). 

In a scenario with a 5 to 15 $\rm M_J$ giant planet, the gravitational torque is much larger than the viscous torque and, thus, the radial location of the trapping is not significantly affected by the disk viscosity \citep[considering typical $\alpha$ parameters of $10^{-4} - 10^{-2}$,][]{Pinilla2012}. 

Finally, we argue that the estimate of this section is valid only on the assumption that a single planet is capable of clearing the observed cavity. We did not explore the scenario of multiple planets. However, such a system seems in contradiction with the presence of diffuse gas inside the cavity (see Sect.\,\ref{Discussion_implications}).

\begin{figure}
   \centering
   \includegraphics[width=9cm]{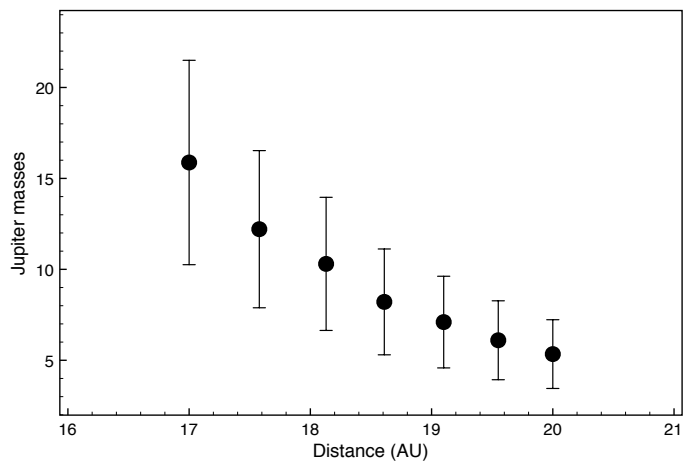}
               \caption{Locations and masses of a {potential} single companion responsible for the observed inner cavity in the disk (see Sect.\,\ref{Discussion_planet} for details). Errors are obtained from the uncertainties in the location of the observed cavity edges \citep[from this work and][]{Brown2009}.  
               }
         \label{Planet}
   \end{figure}

\section{Summary and conclusions}
Transitional disks are considered a key-stage of the process of planet formation. These objects show a large region of highly-depleted dust {surrounded by} a dust rim that can act as a filter, holding back large dust grains but allowing small particles to move inward. Since the particles trapped at the cavity edge can evolve on longer timescales, this scenario has been proposed as a solution to the radial drift barrier. A deeper comprehension of the processes occurring at these cavity edges is, thus, needed. In particular, since small and large grains are expected to behave differently, comparisons of near-IR and millimeter images are an excellent vehicle of investigation. PDI observations of the scattered light from circumstellar disks \citep[e.g.][]{Hashimoto2012, Quanz2013} are providing unprecedented high-contrast near-IR images, whereas SMA \citep{Andrews2009} and ALMA are allowing one to study the thermal emission from an ever growing number of objects. 

The {F4} young star SAO 206462 (HD 135344B) is known to host a transitional disk that shows a narrow dust inner belt \citep{Fedele2008}, a large dust cavity at millimeter wavelengths \citep[$\sim 39$ AU,][]{Brown2009}, and a prominent spiral arms structure \citep{Muto2012}. No cavity was revealed at near-IR wavelengths down to $\sim 28$ AU \citep{Muto2012}. This discrepancy was highlighted for a number of other objects \citep[e.g.][]{Tanii2012, Follette2013}. To explain both this property and the observed spectral deficit at near-IR wavelengths, a scenario with dust filtration and grain growth inside the cavity was proposed \citep{Dong2012}. 

In this paper, we have shown new PDI observations of SAO 206462 obtained with VLT/NACO and tracing the polarized light from the disk as down as $0.1\arcsec$. The main results from our PDI observations are summarized as follows:
\begin{itemize}

\item The disk is detected in polarized light out to $\sim$ 130 AU in both $H$ and $K_s$ band. No particular difference is found between the two bands. The azimuthally-averaged polarized brightness profile is fitted by a power-law with $\beta = -2.9$. However, a broken power-law ($\beta = -1.9$ fitting the inner $\sim 60$ AU, $\beta = -6.0$ fitting outside) provides a better fit.

\item The images reveal the double spiral arms resolved by \citet{Muto2012}, a general dip along the axis at $\theta = 325^{\circ}$ (due to depolarization), and a diffuse enhancement of unclear origin to the east side. 

\item We resolve an inner cavity (brightness down to $2\%$ of a continuous distribution) surrounded by a non-resolved bright rim located at $28 \pm 6$ AU. The rim does not show signs of eccentricity but the azimuthal brightness profile reveals two small bumps, one of which is associated to a spiral arm.    

\end{itemize}

The presence of an inner cavity for $\mu$m-size dust grains constitutes a qualitative explanation for the observed spectral deficit at near-IR wavelengths. Furthermore, a scenario with small particles continuously distributed in the inner disk and subject to intense growth is ruled out. However, the large difference between the cavity size of small ($R \sim 28$ AU) and large ($R \sim 39$ AU) dust grains draws attention to some tantalizing aspects. This cavity is likely to be produced by tidal interaction with one or more companions. In particular, any other known mechanism (photoevaporation, dust grain growth, and MRI) contradicts at least one observational piece of evidence (i.e.\ high accretion rate, inner dust belt, abrupt radial profile at the cavity edge).

The scenario with planet(s) carving out the cavity also provides an explanation for the observed diversity of small and large grain distribution. \citet{Pinilla2012} suggested that the pressure bump induced by an orbiting planet (and trapping large particles only) can be located at radii much larger than the outer edge of the gaseous cavity opened by the same planet. Inspired by this model, we propose a scenario where one giant planet is responsible for holding millimeter-size particles at 39 AU but permits $\mu$m-size grains, perfectly coupled to the gas, to move inward down to 28 AU, where the gas itself is retained by the planet. The presence of gas inside the cavity \citep{Pontoppidan2008} seems to suggest that this is carved out in a narrow ring, rather than in the entire cavity. In any {case}, the suggested scenario implies a depletion of gas in the inner few tens of AU, in agreement with the observations by \citet{Lyo2011}. In particular, we analytically calculate that all observations are consistent with a {potential} planet located at $17-20$ AU with mass $5-15 \ {\rm M_J}$ carving out the cavity. A multiple-planet system is not ruled out by our observations but appears as a contradiction to the presence of diffuse gas inside the cavity.

To fully comprehend the dynamics of dust in the exciting inner regions ($\sim 10-50$ AU) of transitional disks, {complementary observations tracing the distribution of small and large grains are necessary.} In the upcoming era of ALMA {and VLT/SPHERE}, a large sample of high-resolution images of these disks will be available, {both at (sub-)millimeter and optical/near-IR wavelengths}. PDI remains one of the best {techniques} to image circumstellar disks {in the near-IR} as close as $0.1\arcsec$ from the star. Therefore, further PDI investigations of Herbig Ae/Be {and T Tauri stars} and consequent comparisons with millimeter images are crucial. Finally, {L' band images in angular differential imaging mode of SAO 206462 are needed to test the speculated existence of the giant planet sculpting the disk.}   

      \begin{acknowledgements}
      The authors acknowledge the staff at VLT for their excellent support during the observations. We also thank the anonymous referee, A.\,Carmona, and J.\,Pineda for valuable comments on the paper, J.\,Brown for sharing the SMA data, as well as I.\,Kamp and K.\,Pontoppidan for useful discussions. This work is supported by the Swiss National Science Foundation. F.M. is supported by the ETH Zurich Postdoctoral Fellowship Program as well as by the Marie Curie Actions for People COFUND program. This research has made use of the SIMBAD database, operated at CDS, Strasbourg, France.
      \end{acknowledgements}

\bibliographystyle{aa} 
\bibliography{Reference.bib}

\end{document}